
%
%
\def\unredoffs{} \def\redoffs{\voffset=-.31truein\hoffset=-.59truein}
\def\speclscape{}
%
%
%
%
\newbox\leftpage \newdimen\fullhsize \newdimen\hstitle \newdimen\hsbody
\tolerance=1000\hfuzz=2pt
\catcode`\@=11 
\def\bigans{b }
\def\answ{b }
%

\ifx\answ\bigans\message{(This will come out unreduced.}
\magnification=1200\unredoffs\baselineskip=.33truein plus 2pt minus 1pt
\hsbody=\hsize \hstitle=\hsize 
\else\message{(This will be reduced.} \let\l@r=L
\magnification=1000\baselineskip=16pt plus 2pt minus 1pt \vsize=7truein
\redoffs \hstitle=8truein\hsbody=4.75truein\fullhsize=10truein\hsize=\hsbody
\output={\ifnum\pageno=0 
  \shipout\vbox{\speclscape{\hsize\fullhsize\makeheadline}
   \hbox to \fullhsize{\hfill\pagebody\hfill}}\advancepageno
  \else
 \almostshipout{\leftline{\vbox{\pagebody\makefootline}}}\advancepageno
  \fi}
\def\almostshipout#1{\if L\l@r \count1=1 \message{[\the\count0.\the\count1]}
      \global\setbox\leftpage=#1 \global\let\l@r=R
 \else \count1=2
  \shipout\vbox{\speclscape{\hsize\fullhsize\makeheadline}
      \hbox to\fullhsize{\box\leftpage\hfil#1}}  \global\let\l@r=L\fi}
\fi
%
\newcount\yearltd\yearltd=\year\advance\yearltd by -1900

\def\Title#1#2{\nopagenumbers\abstractfont\hsize=\hstitle\rightline{#1}%
\vskip 1in\centerline{\titlefont #2}\abstractfont\vskip .5in\pageno=0}
\def\Date#1{\vfill\leftline{#1}\tenpoint\supereject\global\hsize=\hsbody%
\footline={\hss\tenrm\folio\hss}}
%

\def\draftmode{\message{ DRAFTMODE }\def\draftdate{{\rm preliminary draft:
\number\month/\number\day/\number\yearltd\ \ \hourmin}}%
\headline={\hfil\draftdate}\writelabels\baselineskip=20pt plus 2pt minus 2pt
 {\count255=\time\divide\count255 by 60 \xdef\hourmin{\number\count255}
  \multiply\count255 by-60\advance\count255 by\time
  \xdef\hourmin{\hourmin:\ifnum\count255<10 0\fi\the\count255}}}
\def\nolabels{\def\wrlabeL##1{}\def\eqlabeL##1{}\def\reflabeL##1{}}
\def\writelabels{\def\wrlabeL##1{\leavevmode\vadjust{\rlap{\smash%
{\line{{\escapechar=` \hfill\rlap{\sevenrm\hskip.03in\string##1}}}}}}}%
\def\eqlabeL##1{{\escapechar-1\rlap{\sevenrm\hskip.05in\string##1}}}%
\def\reflabeL##1{\noexpand\llap{\noexpand\sevenrm\string\string\string##1}}}
\nolabels
%
\global\newcount\secno \global\secno=0
\global\newcount\meqno \global\meqno=1
\def\newsec#1{\global\advance\secno by1\message{(\the\secno. #1)}
\global\subsecno=0\eqnres@t\noindent{\bf\the\secno. #1}
\writetoca{{\secsym} {#1}}\par\nobreak\medskip\nobreak}
\def\eqnres@t{\xdef\secsym{\the\secno.}\global\meqno=1\bigbreak\bigskip}
\def\sequentialequations{\def\eqnres@t{\bigbreak}}\xdef\secsym{}
\global\newcount\subsecno \global\subsecno=0
\def\subsec#1{\global\advance\subsecno by1\message{(\secsym\the\subsecno. #1)}
\ifnum\lastpenalty>9000\else\bigbreak\fi
\noindent{\it\secsym\the\subsecno. #1}\writetoca{\string\quad
{\secsym\the\subsecno.} {#1}}\par\nobreak\medskip\nobreak}
\def\appendix#1#2{\global\meqno=1\global\subsecno=0\xdef\secsym{\hbox{#1.}}
\bigbreak\bigskip\noindent{\bf Appendix #1. #2}\message{(#1. #2)}
\writetoca{Appendix {#1.} {#2}}\par\nobreak\medskip\nobreak}
%
%
\def\eqnn#1{\xdef #1{(\secsym\the\meqno)}\writedef{#1\leftbracket#1}%
\global\advance\meqno by1\wrlabeL#1}
\def\eqna#1{\xdef #1##1{\hbox{$(\secsym\the\meqno##1)$}}
\writedef{#1\numbersign1\leftbracket#1{\numbersign1}}%
\global\advance\meqno by1\wrlabeL{#1$\{\}$}}
\def\eqn#1#2{\xdef #1{(\secsym\the\meqno)}\writedef{#1\leftbracket#1}%
\global\advance\meqno by1$$#2\eqno#1\eqlabeL#1$$}
%
\newskip\footskip\footskip14pt plus 1pt minus 1pt 
\def\footnotefont{\ninepoint}\def\f@t#1{\footnotefont #1\@foot}
\def\f@@t{\baselineskip\footskip\bgroup\footnotefont\aftergroup\@foot\let\next}
\setbox\strutbox=\hbox{\vrule height9.5pt depth4.5pt width0pt}
\global\newcount\ftno \global\ftno=0
\def\foot{\global\advance\ftno by1\footnote{$^{\the\ftno}$}}
%
\newwrite\ftfile
\def\footend{\def\foot{\global\advance\ftno by1\chardef\wfile=\ftfile
$^{\the\ftno}$\ifnum\ftno=1\immediate\openout\ftfile=foots.tmp\fi%
\immediate\write\ftfile{\noexpand\smallskip%
\noexpand\item{f\the\ftno:\ }\pctsign}\findarg}%
\def\footatend{\vfill\eject\immediate\closeout\ftfile{\parindent=20pt
\centerline{\bf Footnotes}\nobreak\bigskip\input foots.tmp }}}
\def\footatend{}
%
%
\global\newcount\refno \global\refno=1
\newwrite\rfile
\def\ref{[\the\refno]\nref}
\def\nref#1{\xdef#1{[\the\refno]}\writedef{#1\leftbracket#1}%
\ifnum\refno=1\immediate\openout\rfile=refs.tmp\fi
\global\advance\refno by1\chardef\wfile=\rfile\immediate
\write\rfile{\noexpand\item{#1\ }\reflabeL{#1\hskip.31in}\pctsign}\findarg}
\def\findarg#1#{\begingroup\obeylines\newlinechar=`\^^M\pass@rg}
{\obeylines\gdef\pass@rg#1{\writ@line\relax #1^^M\hbox{}^^M}%
\gdef\writ@line#1^^M{\expandafter\toks0\expandafter{\striprel@x #1}%
\edef\next{\the\toks0}\ifx\next\em@rk\let\next=\endgroup\else\ifx\next\empty%
\else\immediate\write\wfile{\the\toks0}\fi\let\next=\writ@line\fi\next\relax}}
\def\striprel@x#1{} \def\em@rk{\hbox{}}
\def\lref{\begingroup\obeylines\lr@f}
\def\lr@f#1#2{\gdef#1{\ref#1{#2}}\endgroup\unskip}

\def\addref#1{\immediate\write\rfile{\noexpand\item{}#1}} 
\def\startrefs#1{\immediate\openout\rfile=refs.tmp\refno=#1}
\def\xref{\expandafter\xr@f}\def\xr@f[#1]{#1}
\def\refs#1{\count255=1[\r@fs #1{\hbox{}}]}
\def\r@fs#1{\ifx\und@fined#1\message{reflabel \string#1 is undefined.}%
\nref#1{need to supply reference \string#1.}\fi%
\vphantom{\hphantom{#1}}\edef\next{#1}\ifx\next\em@rk\def\next{}%
\else\ifx\next#1\ifodd\count255\relax\xref#1\count255=0\fi%
\else#1\count255=1\fi\let\next=\r@fs\fi\next}
%

%
\newwrite\ffile\global\newcount\figno \global\figno=1
\def\fig{Figure~\the\figno\nfig}
\def\nfig#1{\xdef#1{Figure~\the\figno}%
\writedef{#1\leftbracket fig.\noexpand~\the\figno}%
\ifnum\figno=1\immediate\openout\ffile=figs.tmp\fi\chardef\wfile=\ffile%
\immediate\write\ffile{\noexpand\medskip\noexpand\item{Fig.\ \the\figno. }
\reflabeL{#1\hskip.55in}\pctsign}\global\advance\figno by1\findarg}
\def\xfig{\expandafter\xf@g}\def\xf@g fig.\penalty\@M\ {}
\def\figs#1{figs.~\f@gs #1{\hbox{}}}
\def\f@gs#1{\edef\next{#1}\ifx\next\em@rk\def\next{}\else
\ifx\next#1\xfig #1\else#1\fi\let\next=\f@gs\fi\next}
\newwrite\lfile
{\escapechar-1\xdef\pctsign{\string\%}\xdef\leftbracket{\string\{}
\xdef\rightbracket{\string\}}\xdef\numbersign{\string\#}}

\def\writestop{\def\writestoppt{\immediate\write\lfile{\string\pageno%
\the\pageno\string\startrefs\leftbracket\the\refno\rightbracket%
\string\def\string\secsym\leftbracket\secsym\rightbracket%
\string\secno\the\secno\string\meqno\the\meqno}\immediate\closeout\lfile}}
\def\writestoppt{}\def\writedef#1{}
\def\seclab#1{\xdef #1{\the\secno}\writedef{#1\leftbracket#1}\wrlabeL{#1=#1}}
\def\subseclab#1{\xdef #1{\secsym\the\subsecno}%
\writedef{#1\leftbracket#1}\wrlabeL{#1=#1}}
\newwrite\tfile \def\writetoca#1{}
\def\leaderfill{\leaders\hbox to 1em{\hss.\hss}\hfill}
\def\writetoc{\immediate\openout\tfile=toc.tmp
   \def\writetoca##1{{\edef\next{\write\tfile{\noindent ##1
   \string\leaderfill {\noexpand\number\pageno} \par}}\next}}}

%
\catcode`\@=12 
%
\edef\tfontsize{\ifx\answ\bigans scaled\magstep3\else scaled\magstep4\fi}
\font\titlerm=cmr10 \tfontsize \font\titlerms=cmr7 \tfontsize
\font\titlermss=cmr5 \tfontsize \font\titlei=cmmi10 \tfontsize
\font\titleis=cmmi7 \tfontsize \font\titleiss=cmmi5 \tfontsize
\font\titlesy=cmsy10 \tfontsize \font\titlesys=cmsy7 \tfontsize
\font\titlesyss=cmsy5 \tfontsize \font\titleit=cmti10 \tfontsize
\skewchar\titlei='177 \skewchar\titleis='177 \skewchar\titleiss='177
\skewchar\titlesy='60 \skewchar\titlesys='60 \skewchar\titlesyss='60
\def\titlefont{\def\rm{\fam0\titlerm}
\textfont0=\titlerm \scriptfont0=\titlerms \scriptscriptfont0=\titlermss
\textfont1=\titlei \scriptfont1=\titleis \scriptscriptfont1=\titleiss
\textfont2=\titlesy \scriptfont2=\titlesys \scriptscriptfont2=\titlesyss
\textfont\itfam=\titleit \def\it{\fam\itfam\titleit}\rm}
 \ifx\answ\bigans\else scaled\magstep1\fi
\ifx\answ\bigans\def\abstractfont{\tenpoint}\else
\font\abssl=cmsl10 scaled \magstep1
\font\absrm=cmr10 scaled\magstep1 \font\absrms=cmr7 scaled\magstep1
\font\absrmss=cmr5 scaled\magstep1 \font\absi=cmmi10 scaled\magstep1
\font\absis=cmmi7 scaled\magstep1 \font\absiss=cmmi5 scaled\magstep1
\font\abssy=cmsy10 scaled\magstep1 \font\abssys=cmsy7 scaled\magstep1
\font\abssyss=cmsy5 scaled\magstep1 \font\absbf=cmbx10 scaled\magstep1
\skewchar\absi='177 \skewchar\absis='177 \skewchar\absiss='177
\skewchar\abssy='60 \skewchar\abssys='60 \skewchar\abssyss='60
\def\abstractfont{\def\rm{\fam0\absrm}
\textfont0=\absrm \scriptfont0=\absrms \scriptscriptfont0=\absrmss
\textfont1=\absi \scriptfont1=\absis \scriptscriptfont1=\absiss
\textfont2=\abssy \scriptfont2=\abssys \scriptscriptfont2=\abssyss
\textfont\itfam=\bigit \def\it{\fam\itfam\bigit}\def\footnotefont{\tenpoint}%
\textfont\slfam=\abssl \def\sl{\fam\slfam\abssl}%
\textfont\bffam=\absbf \def\bf{\fam\bffam\absbf}\rm}\fi
\def\tenpoint{\def\rm{\fam0\tenrm}
\textfont0=\tenrm \scriptfont0=\sevenrm \scriptscriptfont0=\fiverm
\textfont1=\teni  \scriptfont1=\seveni  \scriptscriptfont1=\fivei
\textfont2=\tensy \scriptfont2=\sevensy \scriptscriptfont2=\fivesy
\textfont\itfam=\tenit \def\it{\fam\itfam\tenit}\def\footnotefont{\ninepoint}%
\textfont\bffam=\tenbf \def\bf{\fam\bffam\tenbf}\def\sl{\fam\slfam\tensl}\rm}
\font\ninerm=cmr9 \font\sixrm=cmr6 \font\ninei=cmmi9 \font\sixi=cmmi6
\font\ninesy=cmsy9 \font\sixsy=cmsy6 \font\ninebf=cmbx9
\font\nineit=cmti9 \font\ninesl=cmsl9 \skewchar\ninei='177
\skewchar\sixi='177 \skewchar\ninesy='60 \skewchar\sixsy='60
\def\ninepoint{\def\rm{\fam0\ninerm}
\textfont0=\ninerm \scriptfont0=\sixrm \scriptscriptfont0=\fiverm
\textfont1=\ninei \scriptfont1=\sixi \scriptscriptfont1=\fivei
\textfont2=\ninesy \scriptfont2=\sixsy \scriptscriptfont2=\fivesy
\textfont\itfam=\ninei \def\it{\fam\itfam\nineit}\def\sl{\fam\slfam\ninesl}%
\textfont\bffam=\ninebf \def\bf{\fam\bffam\ninebf}\rm}
%
%

\hyphenation{anom-aly anom-alies coun-ter-term coun-ter-terms}
\def\inv{^{\raise.15ex\hbox{${\scriptscriptstyle -}$}\kern-.05em 1}}

\def\Dsl{\,\raise.15ex\hbox{/}\mkern-13.5mu D} 
\def\dsl{\raise.15ex\hbox{/}\kern-.57em\partial}

\font\bigit=cmti10 scaled \magstep1
\def\lspace{\ifx\answ\bigans{}\else\qquad\fi}
\def\lbspace{\ifx\answ\bigans{}\else\hskip-.2in\fi} 
\def\boxeqn#1{\vcenter{\vbox{\hrule\hbox{\vrule\kern3pt\vbox{\kern3pt
	\hbox{${\displaystyle #1}$}\kern3pt}\kern3pt\vrule}\hrule}}}
\def\mbox#1#2{\vcenter{\hrule \hbox{\vrule height#2in
		\kern#1in \vrule} \hrule}}  
%

\def\darr#1{\raise1.5ex\hbox{$\leftrightarrow$}\mkern-16.5mu #1}

\def\half{{\textstyle{1\over2}}} 
\def\roughly#1{\raise.3ex\hbox{$#1$\kern-.75em\lower1ex\hbox{$\sim$}}}

\def\perpp{{\scriptscriptstyle\perp}}

\def\half{{1\over 2}}

\def\kbT{k_{\scriptscriptstyle\rm B}T}

\def\bold#1{\setbox0=\hbox{$#1$}%
     \kern-.010em\copy0\kern-\wd0
     \kern.025em\copy0\kern-\wd0
     \kern-.020em\raise.0200em\box0 }

\def\dz{\partial_z}

\def\ts{\theta_6}

\def\cross{\!\times\!}
\def\dnb{\delta {\vec n}}

\lref\BOUii{
F.~Livolant
and Y.~Bouligand, J. Phys. (Paris) {\bf 47} 1813 (1986); For tilt and moir\'e
boundaries in the non-biological polymer PBZO (poly-paraphenylene
benzobisoxazole) see
D.C.~Martin and E.L.~Thomas, Phil. Mag. A {\bf 64}, 903 (1991).
}
\lref\TGB{
S.R.~Renn and T.C.~Lubensky, Phys. Rev. A {\bf 38}, 2132 (1988);  For
experiments, see
G.~Srajer, R.~Pindak, M.A.~Waugh and
J.W.~Goodby, Phys. Rev. Lett. {\bf 64}, 1545 (1990); see also
P.G.~de Gennes, Solid State Commun. {\bf 10}, 753 (1972).
}\lref\DGP{
P.G.~de Gennes and J.~Prost, The Physics of Liquid Crystals, Second
ed.,
(Oxford University Press, New York, 1993).
}\lref\KN{
For a more detailed discussion, see
R.D.~Kamien and D.R.~Nelson, Institute for Advanced Study Preprint
IASSNS-HEP-94/68 and
in preparation.}
\lref\IND{
V.L.~Indenbom and A.N.~Orlov, Usp. Fiz. Nauk {\bf 76} 557 (1962) [
Sov. Phys. Uspekhi {\bf 5} 272 (1962)].
}\lref\MN{
M.C.~Marchetti and D.R.~Nelson, Phys. Rev. B {\bf 41} 1910 (1990).
}\lref\KO{
M.~Kl\'eman and P.~Oswald, J. Phys. (Paris) {\bf 43} 655 (1982).
}\lref\BOL{
W.A.~Bollman, Crystal Defects and Crystalline Interfaces, (Springer-Verlag,
Berlin, 1970);
J.P.~Hirth and J.~Lothe, Theory of Dislocations, Second ed. (Wiley, New York,
1982).
}\lref\SWM{
D.C.~Wright and N.D.~Mermin, Rev. Mod. Phys. {\bf 61}, 385 (1989);
see also M. Kl\'eman, J. Phys. (Paris) {\bf 46}, 1193 (1985).
}

\Title
{IASSNS-HEP-94/33}
{Polymer Braids and Iterated Moir\'e Maps\footnote{}
{To Appear in the {\sl Proceedings of the Wiener 1994 Centennial Symposium}
edited
by D. Jerison, I.M. Singer and D.W. Strock }}

\centerline{David R. Nelson}
\centerline{\sl Lyman Laboratory of Physics,
Harvard University, Cambridge, MA 02138}\smallskip\centerline{and}
\smallskip
\centerline{Randall D. Kamien}
\centerline{\sl School of Natural Sciences, Institute for Advanced Study,
Princeton, NJ
08540}
\vskip .3in
Crystalline order in dense packings of long polymers with a definite handedness
is
difficult to reconcile with the tendency of these chiral objects to twist and
braid about each other.  If the chirality is weak, the state of lowest energy
is
a triangular lattice of rigid rods.
When the chirality is strong, however,
screw dislocations proliferate, leading to either a tilt grain boundary phase
or a new
``moir\'e state'' with twisted bond order. In the latter case, polymer
trajectories in the plane perpendicular to their average direction are
described by
iterated moir\'e maps of remarkable complexity, reminiscent of dynamical
systems.
\Date{28 December 1994}
\newsec{Introduction}
A notable feature of biological materials is the profusion of long polymer
molecules
with a definite handedness.  DNA, polypeptides (such as
poly-$\gamma$-benzyl-glutamate)
and polysaccharides (such as xanthan) can all be
synthesized with a preferred chirality.
Long polymers in dense solution often crystallize into a hexagonal columnar
phase, {\sl
i.e.}, a lattice of rod-like objects with the cross section a triangular
lattice.
When the polymers
are chiral this close packing into a triangular lattice
competes with the tendency for the polymers to twist macroscopically
\BOUii\
as in cholesteric liquid crystals.  Similar to the twist grain boundary
phase of chiral smectics \TGB , macroscopic chirality can proliferate when
screw
dislocations enter the crystal.  Though the screw dislocations cost a finite
energy, when the energy increase is offset by the decrease in free energy from
twisting, the
screw dislocations will penetrate the sample.
If the chirality is weak, a defect
free hexagonal columnar phase persists.

In this note we discuss the effect of chirality
on the hexagonal columnar phases \DGP\ of long
polymers \KN .
We neglect for simplicity heterogeneity along the polymer backbones and work
with a two component displacement field perpendicular to the local polymer
direction.
The
usual chiral term relevant for cholesteric liquid crystals produces a polymer
tilt grain
boundary phase,
similar to the
smectic-$A^*$ phase \TGB .  We find, as well, an additional
term in the free energy which favors the rotation of the bond order along the
average
polymer direction.
This term leads to braided polymers with twisting describable
by a sequence of moir\'e patterns.  The polymer braids can be described by
a map from one perfect crystalline lattice to another perfect lattice rotated
by a relative angle $\phi$ and separated by honeycomb network of defects.
When $\phi$ takes on certain lock-in angles, the
free energy of the crystal distortion will go through a minimum.  The resulting
map creates moir\'e patterns when the monomer locations are projected down the
polymer
axes.  Upon iteration, corresponding to many periodically spaced defect walls,
these moir\'e maps exhibit complex, self-similar behavior, reminiscent of
chaotic
dynamical systems.  Little appears to be known about the properties of these
iterated
moir\'e maps.

In the next section we review the continuum elastic theory which applies to a
polymer crystal.  We
show how the energy of dislocation defects in the crystal determines
the critical values of the chiral coupling constants
at which chirality first enters.  We describe
the smectic twist grain boundary state as well as its polymer analogue, the
tilt
grain boundary state.  In the last section we describe a microscopic model of
the
moir\'e phases, and discuss the lock-in angles which will produce moir\'e maps
and
braided polymer crystals.  Finally, in the appendix we prove that any rotation
angle of a crystal
which produces
a non-trivial coincidence lattice (as do the special lock-in angles discussed
above)
is an irrational fraction
of $2\pi$.

\newsec{Statistical Mechanical Model and the Tilt Grain Boundary Phase}

In order to calculate averages of physical quantities, we integrate over
the ensemble of all states, with each state weighted by the Boltzmann
weight
\eqn\ebolt{
P[\vec u] = {\cal Z}^{-1} \exp\{-F[\vec u]/\kbT\}
}
where $\vec u(x_1,x_2,x_3)$ is the two-dimensional displacement field of the
polymers from
a perfect triangular lattice of rods, and $\cal Z$ is the partition function.
The elastic free energy, appropriate
for a crystalline state of polymers lying on average along the $z$-axis, is
\eqn\efreelast{
F[\vec u]=\int d^3\!x\,\left\{\mu u_{ij}^2 + {\lambda\over 2}u_{ii}^2 +
K_3(\partial_z^2u_i)^2 -\gamma\nabla_{\!\perpp}\cross\dnb
-\gamma'\partial_z\ts\right\}
}
where $u_{ij} =\half(\partial_iu_j+\partial_ju_i)$, $\delta n_i=\partial_z
u_i$,
and $\ts=\half\epsilon_{\mu\nu\rho}n_\mu\partial_\nu u_\rho
=\half\epsilon_{ij}\partial_iu_j$ in the nematic ($\vec n_0 =\hat z$) ground
state.
These fields measure respectively the local strain,
tilt and bond order of the crystal.  Here and throughout Roman indices
run from $1$ to $2$ while Greek indices run from $1$ to $3$.
Terms which are not invariant under spatial reflection are chiral and are not
allowed in systems with mirror symmetry.  The last two terms in \efreelast\ are
the
only allowed couplings in a chiral polymer system to lowest order in the
gradients of
$\vec u$.  The term proportional to
$\nabla_{\!\perpp}\cross\dnb$ favors a cholesteric-like twisting \DGP\ of the
local
polymer direction along a pitch axis, perpendicular to the planes in which
the polymers, on average, lie.  The term proportional to $\dz\ts$ favors
a twisting of the bond order along the polymer direction.  This term, and
its effect on the polymer trajectories, can be described by moir\'e maps.
In the columnar crystal the
two chiral terms are the same if $\partial_z\partial_i=\partial_i\partial_z$.
However, in the presence of crystal dislocations derivatives do {\sl not}
commute,
$\nabla_{\!\perpp}\cross\dnb = \epsilon_{ij}\partial_i\partial_z
u_j\ne\partial_z
\epsilon_{ij}\partial_iu_j = \partial_z\ts$.  A dislocation line is
characterized by
the value of the line integral around this defect \IND
\eqn\bur{
\oint_\Gamma ds_\mu \partial_\mu u_i = -b_i
}
where $\vec b$ is the Burgers vector.  The Burgers vector must be an integer
linear combination
of lattice vectors.  It points parallel to the defect axis for the screw
dislocation shown in Figure 1.

Various arrangements of
dislocations lead to the twisting and braiding of polymers favored by $\gamma$
and $\gamma'$.
Burgers vectors of dislocations in a hexagonal columnar phase
lie in the $xy$-plane, and there are three generic types \DGP :
a screw dislocation, an edge dislocation
with tangent along $\hat z$, and an edge dislocation lying in the
$xy$-plane.  The latter defect requires aligned polymer ends which we can
neglect if
the polymers are very long.  The remaining dislocations must lie in a plane
spanned by their Burgers
vector $\vec b$ and $\hat z$, which amounts to choosing dislocation complexions
with
$\alpha_{xy}=\alpha_{yx}$, where the dislocation density tensor $\alpha_{\gamma
i}$ is the
density of dislocations with tangents
along the $\gamma$-direction with Burgers vectors pointing
in the $i$ direction \MN .

Proceeding as
in \refs{\IND,\MN} we introduce a new field $w_{\gamma i} \equiv
\partial_\gamma u_i$
away from any dislocations.  The non-commutivity of
the derivatives of $\vec u$ is represented by the dislocation density,
$\epsilon_{\mu\nu\gamma}
\partial_\nu
w_{\gamma i} = -\alpha_{\mu i}$.
One can solve for $w_{\gamma i}$ in terms of the dislocation density
$\alpha_{\gamma i}$ and find the equilibrium displacement field in the presence
of
crystal defects \KN . In terms
of the nematic and bond order field non-commutivity of derivatives means
$2\dz\ts-\nabla_{\!\perpp}\cross\dnb =
-{\rm Tr}[\alpha]$.  The energy per unit length $f_{\rm s}$
of a screw dislocation is finite,
while that of an edge dislocation lying along the $\hat z$ direction diverges
logarithmically with system size \KO .  A screw dislocation is illustrated in
Figure 1.
Depending on how the dislocations are combined together a screw dislocation
array can either
produce a non-vanishing $\nabla_{\!\perpp}\cross\dnb$ or $\dz\ts$.  Our
predictions
for the phase diagram as a function of $\gamma$ and $\gamma'$
are summarized in Figure 6.

If the chirality is strong and $\gamma\gg\gamma'$ we expect
the polymer analogue of the Renn-Lubensky
twist-grain-boundary state \TGB\ of chiral smectic liquid crystals. Smectic
liquid
crystals are, in some sense, dual to the hexagonal columnar phases considered
here \DGP .
Smectics are composed of sheets of rod-like molecules;  they are crystalline in
one
direction (the direction perpendicular to the layers) instead of two.  If the
molecules
are chiral these layers want to twist, thus disrupting the crystalline order.
At low
chiralities, macroscopic twist is suppressed.  Above a certain threshold
chirality,
screw dislocations enter the sample, causing the
smectic slabs to undergo, on average, a uniform rotation along a direction
perpendicular
to the layer normals.  In Figure 2
we show the structure of this state.
Polymer crystals, as well, can have a state characterized by a constant overall
twist
perpendicular to the polymer axes.
The state consists of a parallel array of tilt grain boundaries, separated by
$d'$ along
the rotation axis, bounding regions of perfect polymer crystals similar to
the smectic planes in Figure 2.
Each of these tilt grain boundaries (TGB)
is composed of a parallel array of
screw dislocations lying, say,
in the $xz$-plane, pointing along the $x$-axis and uniformly spaced along $\hat
z$ with spacing
$d$.
As illustrated in the upper inset of Figure 6 this dislocation texture causes a
discrete rotation
$\phi=\tan^{-1}(b/d)$ (with $b=a_0$, corresponding to the shortest allowed
Burgers vector)
in
the average polymer direction across the boundary.
The spatial integral of $\nabla_{\!\perpp}\cross\dnb$ is non-zero, while the
integral of $\dz\ts$ vanishes.  The TGB state appears when the
chiral coupling $\gamma$ exceeds the critical value $\gamma_c=f_{\rm s}/b$ \KN
{}.

\newsec{The Moir\'e State}

The second chiral coupling $\gamma'$ has no analogue in chiral smectics.  To
find a
configuration of screw dislocations which exploits this form of chirality
we search for a dislocation texture which produces a displacement
depending only on $z$. The only texture which does
not produce divergent elastic energy is one in which the only
non-zero
components of $\alpha_{\gamma i}$ are $\alpha_{xx}=\alpha_{yy}$ \refs{\MN,\KN}.
A honeycomb array of
screw dislocations, on average, produces this dislocation texture.
We
now find
that for a single such grain boundary $\nabla_{\!\perpp}\cross\dnb$ vanishes
far from
the boundary while
$\partial_z\ts$ does not. This sort of grain boundary thus causes a net twist
of
the hexatic bond order parameter $\ts$ while imposing no net
$\nabla_{\!\perpp}\cross\dnb$.
When $\gamma'\gg\gamma$ and the chirality is large, screw dislocations
penetrate
for $\gamma'> \gamma'_c = 2f_{\rm s}/b$.  By choosing the rotation angle across
the
honeycomb dislocation network to produce a high density of coincidence lattice
sites \BOL
we produce
especially low strain energies across the boundary. The superposition of
triangular polymer
lattices below and above the boundary forms a moir\'e pattern.  The
superposition of many such ``moir\'e sheets'' along the $z$-axis braids
the polymers
with deep minima in the energy at
certain lock-in angles.
Figure 3 illustrates the mapping of polymers across the moir\'e plane.
Polymers
in
the lower half-space (circles) must be connected to the closest available
polymer in the
upper half-space (crosses) to minimize bending energy.
Note that the map has a discrete translational
symmetry, in the sense that any coincidence site could be a center of rotation.
 Especially
simple moir\'e maps arise for the rotation angles
\eqn\elock{
\phi_n = 2\,\tan^{-1}\left[{\sqrt{3}\over 3(2n+1)}\right]
}
$n=1,2,\ldots$.  Figure 3 shows these maps for $n=1,\ldots,4$.
It is shown in Appendix A that all such angles are irrational fractions of
$2\pi$ \KN\ so that the structure never repeats upon iteration.
Around each coincidence point
there are $n$ concentric rings of helical polymers.  The lattice of coincidence
points is
also a triangular lattice, but with a spacing $a_n=a_0\sqrt{1+3(2n+1)^2}/2$,
where $a_0$ is the original lattice constant.
The geometrical origin of such energetically preferred lock-in angles has no
analogue
in chiral smectics.  The exact choice of lock-in angles and spacing between
moir\'e planes
must be settled by detailed energetic calculations.

Upon two iterations of the moir\'e map separating three regions of
polymer crystal, the first coincidence lattice is rotated with respect to
the second coincidence lattice by precisely the angle of rotation $\phi_n$.
Thus the composite coincidence
lattice is the ``coincidence lattice of coincidence lattices'', with lattice
constant $a_n^2/a_0$.   Moir\'e maps iterated $p$ times lead to triangular
composite
coincidence lattices with spacing $a_n(a_n/a_0)^{p-1}$, {\sl i.e.} to
ever sparser lattices of fixed points with intricate fractal
structure in between them.  Figure 4 shows the projected polymer
paths for a lock-in angle of a square lattice (with $\phi_1=\tan^{-1}(3/4)
\approx 36.9^{\circ}$) iterated $p=1,\ldots,4$ times.
For a square lattice an especially simple sequence of lock-in angles is given
by
\eqn\elocsq{
\phi_n^{\rm square}
= \tan^{-1}\left[{2n+1\over 2n(n+1)}\right],
}
leading to coincidence lattice spacings
$a_n^{\rm square}=a_0\sqrt{n^2+(n+1)^2}$.

In contrast to
the
TGB state polymers for $p\gg 1$ are
highly entangled and wander far from straight line trajectories.
The polymer configurations and the
dislocations leading to them are shown in Figure 5 for the $n=1$ moir\'e map of
a
triangular lattice iterated nine times.  Near the
moir\'e planes both $\ts$ and $\nabla_{\!\perpp}\cross\dnb$ are nonzero.
The center polymer is
a fixed point of all the maps.  Any such fixed polymer has
a halo of others twisting around it.  In this special tube, the
nematic order parameter takes on the texture of a double twist cylinder as
found
in the low-chirality limit of
blue phases of cholesteric liquid crystals \SWM
The moir\'e state takes advantage of both double twist energies and the
new chiral coupling $\gamma'$.

The phase diagram in Figure 6 summarizes our conclusions.  As for chiral
smectics
\TGB , it is possible to cast the theory in a form similar to the
Ginzburg-Landau theory of Type II superconductors in a magnetic field \KN .
The couplings $\gamma$ and $\gamma'$ represent two distinct ``magnetic fields''
in
this analogy. The hexagonal columnar
phase expels all macroscopic chirality.  The {\sl two} chiral couplings
$\gamma$ and $\gamma'$ cause screw dislocations to penetrate the crystal above
critical strengths $\gamma_c$ and $\gamma_c'$, similar to the penetration of
Abrikosov
vortices above the lower critical field $H_{c1}$.
The TGB phase predicted here for chiral polymers is similar to that already
observed
experimentally in chiral smectics \TGB .
The braided moir\'e state is qualitatively
new, and its experimental observation would be of considerable interest.
In Figure 7 we show 40 random polymers subject to $99$ iterations of the $n=1$
map for the triangular lattice, all near a central, fixed coincidence lattice
point. Surprising
little is known about the intricate trajectories produced by such iterated
moir\'e maps.
These could be studied experimentally via neutron diffraction in hexagonal
columnar crystals with a dilute concentration of deuterated polymer strands.
Numerical studies of the local fractal dimension and Lyapunov exponent of these
polymer trajectories are currently
in progress.
\newsec{Acknowledgements}
It is a pleasure to acknowledge stimulating conversations with T.~Lubensky,
R.~Meyer,
P.~Taylor, E.~Thomas,
and J.~Toner.
DRN acknowledges the hospitality of Brandeis University, AT\&T Bell
Laboratories, and Exxon Research and Engineering, as well as support from the
Guggenheim
Foundation and the National Science Foundation Grant No. DMR-94-17047,
and in part through the Harvard Materials
Research Science and Engineering Center, under Grant No. DMR-94-17047.
RDK was supported by National Science Foundation Grant
No.~PHY92--45317

\appendix{A}{Irrationality of the Lock-In Angles}
The special lock in angles that we have found for the triangular lattice are
all
irrational fractions of $2\pi$.  We will prove a more general theorem in this
appendix that
applies to any rotation which leads to a coincidence lattice with a finite
lattice
constant.

Let ${\bf R}(\theta,\Lambda)=\Lambda'$ be the
lattice generated by rotating a lattice of points $\Lambda$ by
$\theta$ around a lattice site at the origin.
Consider a regular lattice $\Lambda_1$, a set of ordered pairs $\{
(x_1,y_1),(x_2,y_2),\ldots\}$.
Suppose there is an angle $\theta^*$ for which the intersection of $\Lambda_1$
and ${\bf R}
(\theta^*,\Lambda_1)$ is a lattice (not equal, as in the pathological case
$\Lambda_1=
{\bf R}(\theta^*,\Lambda_1)$, to the original lattice)
with a finite lattice constant.  These angles are
special lock-in angles which keep a finite fraction of polymers from
stretching.
Let $\Lambda_1^n={\bf
R}(n\theta^*,\Lambda_1)$.
Let the intersection of $\Lambda_1^0$ and $\Lambda_1^1$ be the coincidence
lattice
$\Lambda_2^0$, and similarly $\bigcap_{j=m}^{m+k}\Lambda_1^j =
\Lambda_{k+1}^m$.
Note that $\Lambda_k^m={\bf R}(m\theta^*,\Lambda_k^0)$.
The intersection of $\Lambda_1^1$ and $\Lambda_1^2$ is also a coincidence
lattice, but as $\Lambda_1^1$ and $\Lambda_1^2$ are rotated by $\theta^*$ from
$\Lambda_1^0$ and $\Lambda_1^1$, respectively, their intersection,
$\Lambda_2^1$ will
be rotated by $\theta^*$ from $\Lambda_2^0$.  Thus, since the intersection of
$\Lambda_1^0$,
$\Lambda_1^1$ and $\Lambda_1^2$ is just the intersection of $\Lambda_2^0$ and
$\Lambda_2^1$
the resulting coincidence lattice is the coincidence lattice of the coincidence
lattices
$\Lambda_2$.

In general, if we look at $\Lambda_{n+1}^0$
\eqn\inters{
\Lambda_{n+1}^0\equiv\bigcap_{j=0}^n\Lambda_1^j = \left(
\bigcap_{j=0}^{n-1}\Lambda_1^j\right)
\bigcap\left(\bigcap_{j=1}^n\Lambda_1^j\right)\equiv
\Lambda_n^0\bigcap{\bf R}(\theta^*,\Lambda_n^0).
}
Thus $\Lambda_{n+1}^0$ is the coincidence
lattice of two coincidence lattices $\Lambda_n$.  Therefore as we add layer
after layer, consecutively rotating by $\theta^*$ we will produce ever sparser
lattices with growing lattice constants.

Now assume that $\theta^*$ is a rational fraction of $2\pi$, so $\theta^* =
2\pi{p\over q}$.
Then, after $q$ iterations of the rotation the lattices would repeat, or, in
other
words, $\Lambda_1^q=\Lambda_1^0$. This would mean that $\Lambda_{q+1}^0\equiv
\bigcap_{j=0}^{q}\Lambda_1^j=\bigcap_{j=0}^{q-1}\Lambda_1^j\equiv\Lambda_q^0$
since $\Lambda_1^q$ has already been included in the intersection of the
first $q$ lattices.  This contradicts the conclusion of the preceding
paragraph.
Therefore it is impossible to have a rotation angle
$\theta^*$ which is both a rational fraction of $2\pi$ {\sl and} produces
a coincidence lattice with a non-trivial but finite lattice constant.  Thus
all lock-in angles must be irrational fractions of $2\pi$, which is what we
were
to prove.

\nfig\figi{
A single screw dislocation in a polymer crystal.  The dark horizontal
line is the screw dislocation.}

\nfig\figii{
The smectic twist grain boundary state.
Each plane is a smectic layer and each stack
is separated by a smectic Twist Grain Boundary. Figure provided courtesy
of Tom Lubensky.}

\nfig\figiii{
A single moir\'e map for triangular lattices with $n=1,2,3,4$
($\phi_n=$ $21.8^{\circ}$, $13.2^{\circ}$, $9.4^{\circ}$, $7.3^{\circ}$
respectively).  The crosses
can be thought of as the ends of polymers in a perfect crystalline region, as
can
the circles.  They join by connecting to the nearest polymer in order to reduce
the
elastic energy.  The shaded lines are the screw dislocations which make up a
honeycomb
network.}

\nfig\figiv{
The projected top view of a moir\'e map on a square lattice with rotation
angle $\tan^{-1}(3/4)$.  The four boxes show the projected polymer
paths after $p=1,2,3,4$ iterations.}

\nfig\figv{
The moir\'e state.  The thick tubes
running in the $\hat z$ direction are polymers, while the dark lines are
stacked honeycomb arrays of screw dislocations.  The intersection of
these polymers with any constant $z$ cross section away from the hexagonal
defect arrays forms a perfect triangular lattice.}

\nfig\figvi{
Phase diagram of a chiral polymer crystal.  Insets are representative
tilt (TGB) and moir\'e grain boundaries.  Shaded lines are screw dislocations.}

\nfig\figvii{
A projected top view of $40$ random
polymers paths
resulting from the moir\'e map with $n=1$ iterated 99 times.
There is an exceptional fixed point of all 99 maps at the center.}

\footatend\vfill\supereject\immediate\closeout\rfile\writestoppt
\baselineskip=.33truein\centerline{{\bf References}}\bigskip{\frenchspacing%
\parindent=20pt\escapechar=` \input refs.tmp\vfill\eject}\nonfrenchspacing
\vfill\eject\immediate\closeout\ffile{\parindent40pt
\baselineskip.33truein\centerline{{\bf Figure Captions}}\nobreak\medskip
\escapechar=` \input figs.tmp\vfill\eject}

\bye
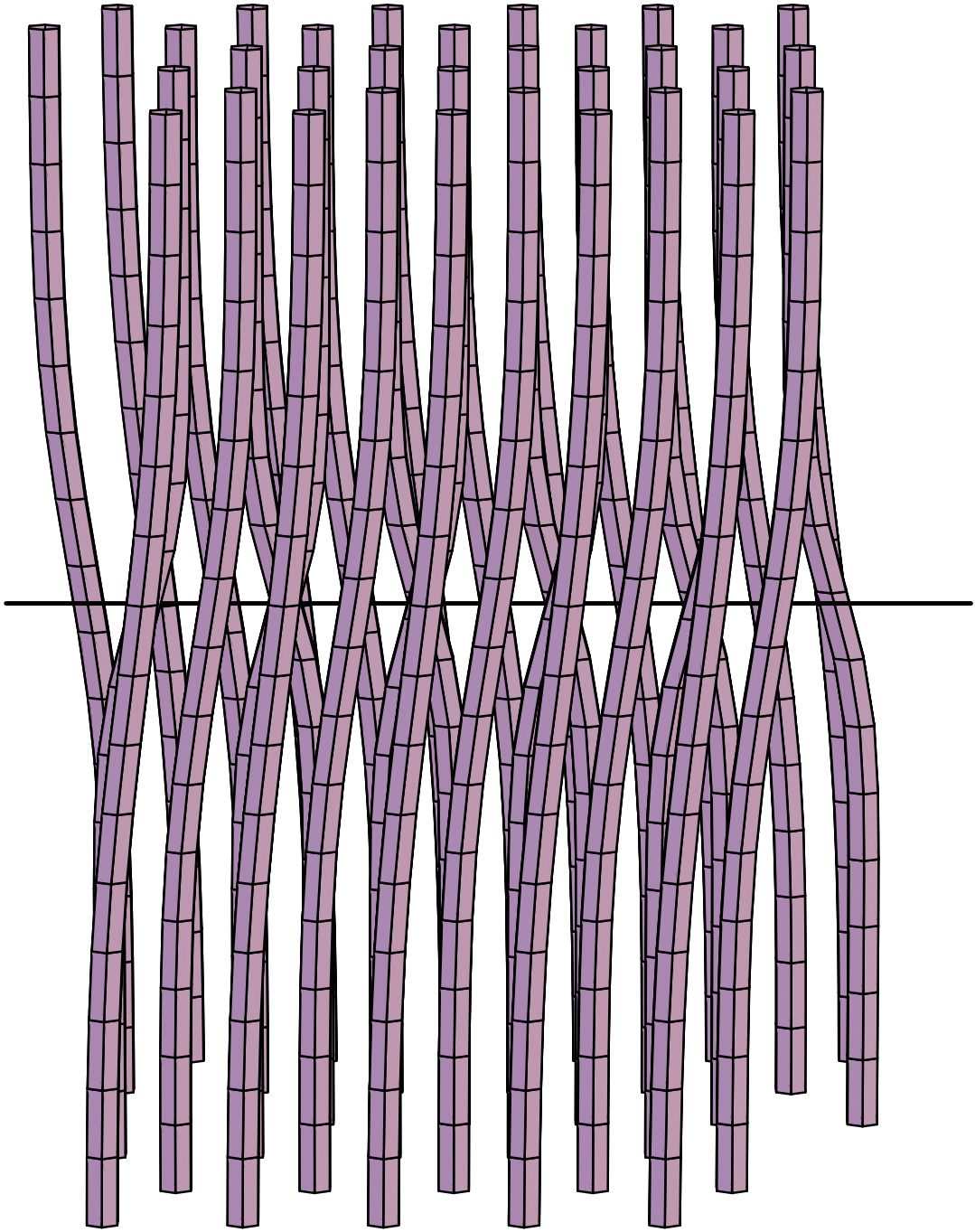
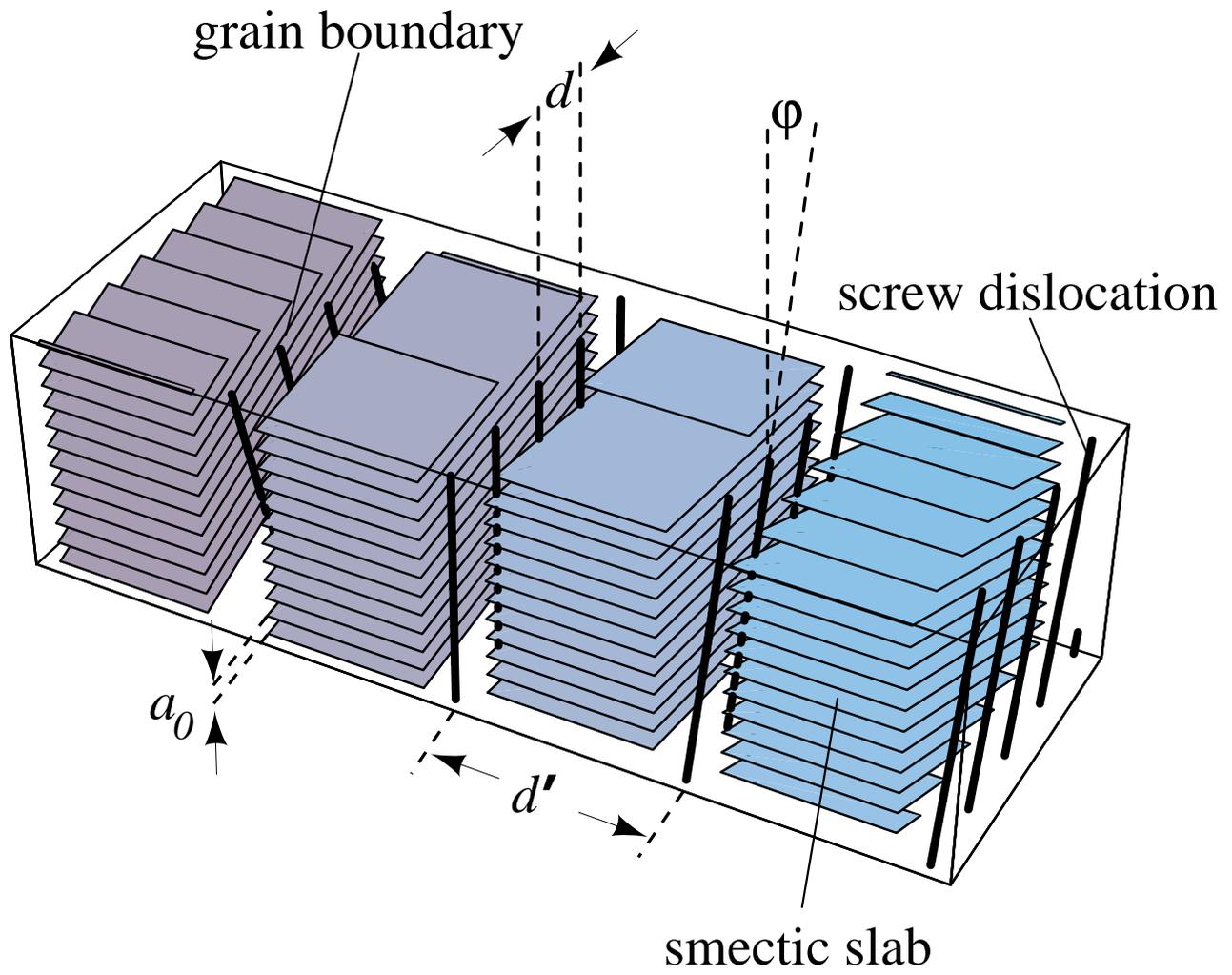
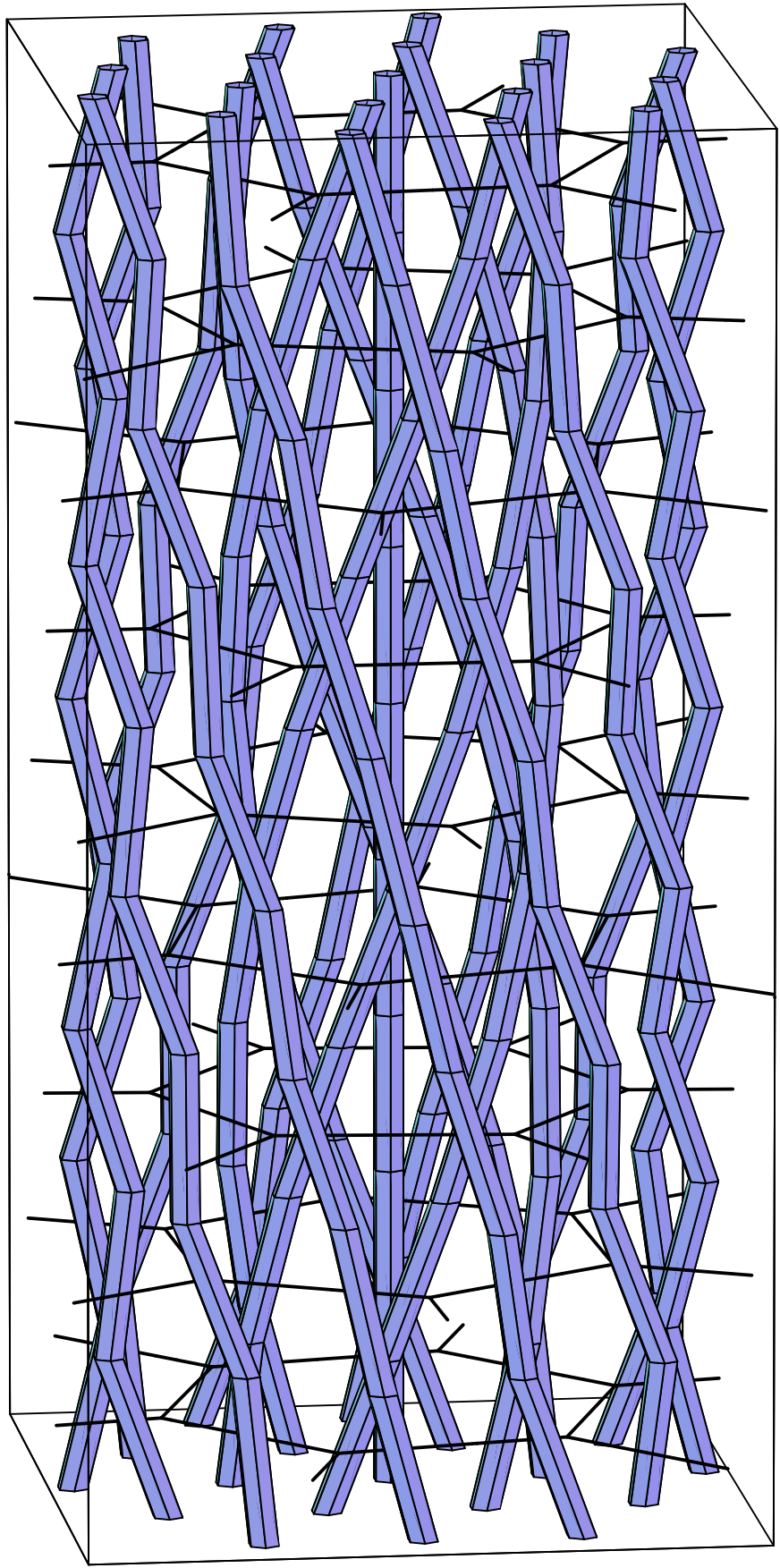
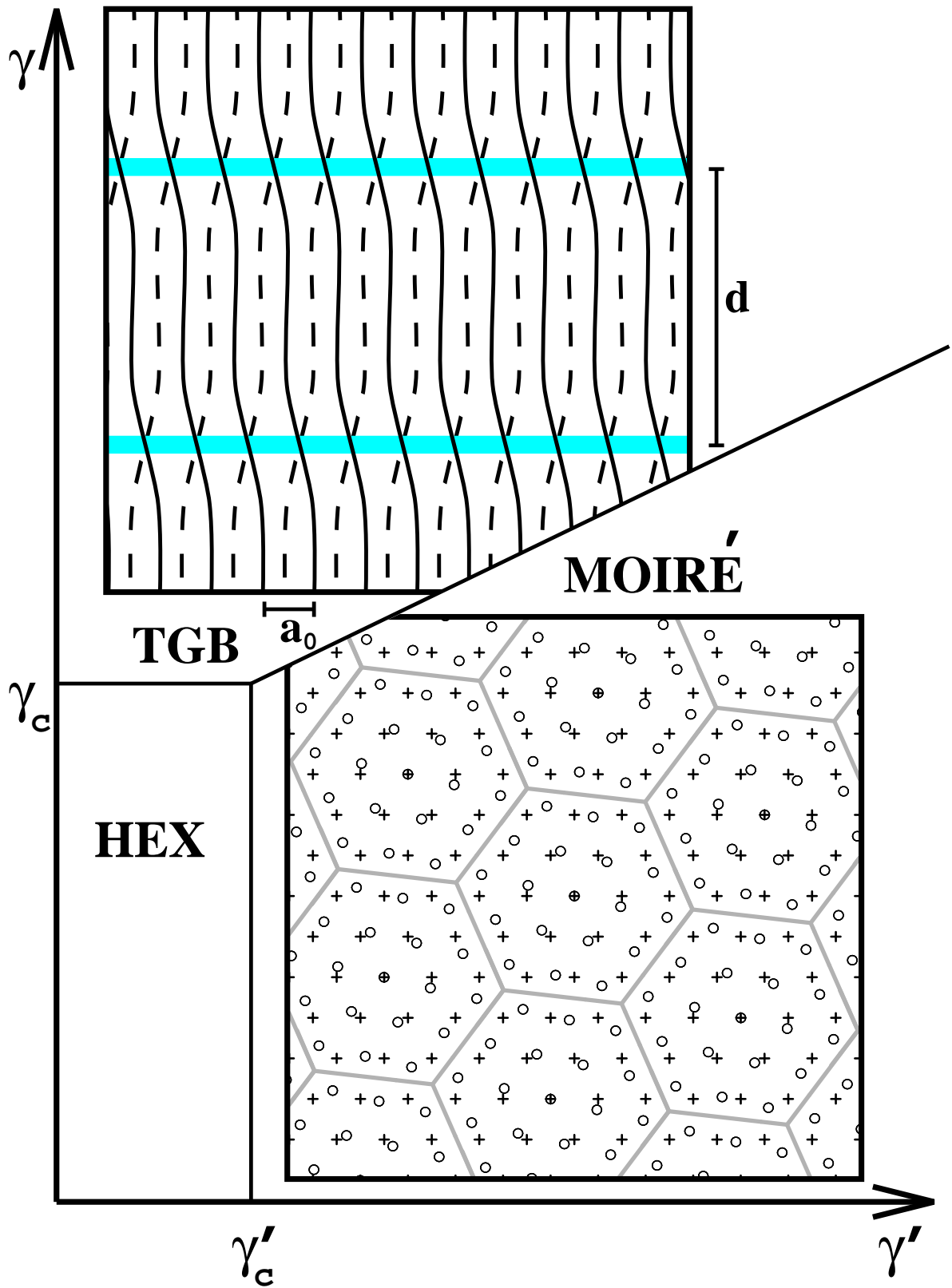
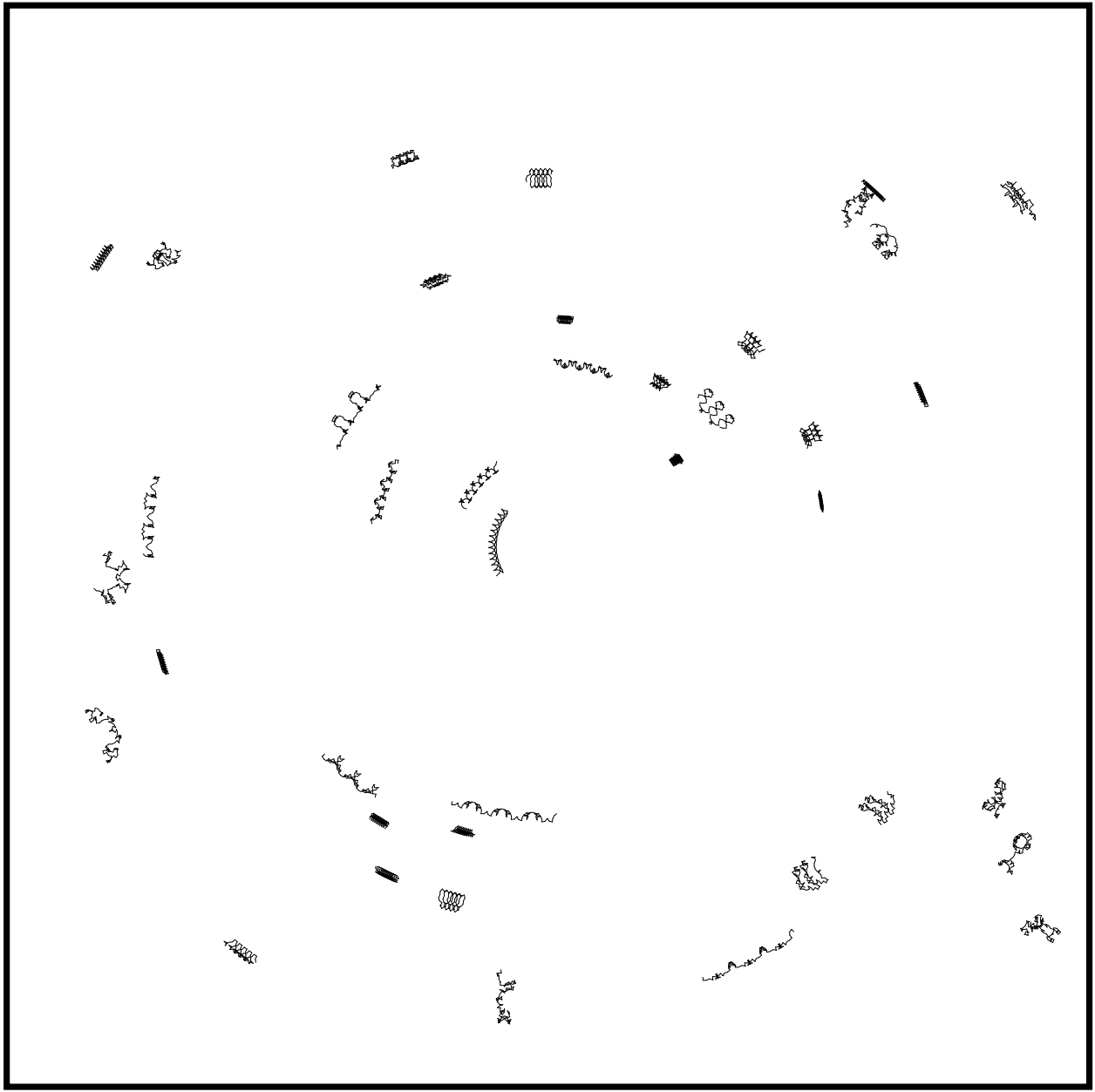
\end